\documentclass[12pt,tightenlines,nofootinbib,prx]{revtex4}
\usepackage[usenames,dvipsnames,svgnames,table]{xcolor}
\usepackage{amssymb}
\usepackage{amsmath}
\usepackage{amsfonts}
\usepackage[stable]{footmisc}
\usepackage{hyperref}
\usepackage{mathrsfs}
\usepackage{bbm}
\usepackage{tikz}
\usetikzlibrary{arrows,fit,positioning,automata}
\usetikzlibrary{decorations,decorations.pathmorphing,decorations.pathreplacing,shapes.arrows,shapes.geometric}

\newenvironment{definition}[1][Definition]{\begin{trivlist}
\item[\hskip \labelsep {\bfseries #1}]}{\end{trivlist}}

\newcommand{\ket}[1]{\left| #1 \right>} 
\let\baraccent=\= 
\renewcommand{\=}[1]{\stackrel{#1}{=}} 

\begin{document}

\title{Contextuality: Wheeler's universal regulating principle}

\author{Ian T. Durham}
\email[]{idurham@anselm.edu}
\affiliation{Department of Physics, Saint Anselm College, Manchester, NH 03102}
\date{\today}

\begin{abstract}
In this essay I develop quantum contextuality as a potential candidate for Wheeler's universal regulating principle, arguing --- \textit{contrary} to Wheeler --- that this ultimately implies that `bit' comes from `it.'
\end{abstract}

\maketitle

\begin{quote}
\textit{All I did this week was rearrange bits on the internet. I had no real impact on the physical world.}
\begin{flushright} --- Dilbert \end{flushright}
\end{quote}

\section{It and Bit}
In his Oersted Medal acceptance address in 1983, John Wheeler expressed his view that the primary task of what he called the ``coming third era of physics'' was the identification of a universal regulating principle arising from a ``regularity based on chaos, of `law without law' '' \cite{Wheeler:1983fk}.  This third era of physics will clearly require a radical reconsideration of many of our most cherished ideas.  Indeed, such reconsiderations have, in recent years, led to the development of categorical quantum mechanics \cite{Abramsky:2004fk,Abramsky:2008uq}, the derivation of physical laws from ordering relations such as posets \cite{Knuth:2003vn,Knuth:2010ys,Coecke:2011uq}, and the introduction of topos theory to theoretical physics \cite{Isham:2010kx,Doring:2011fk}, to name but a few.  In the latter approach, D\"{o}ring and Isham have even attempted to answer Heidegger's amorphous question, `What is a Thing?' \cite{Heidegger:1967fk,Doring:2011fk}.

Indeed, what \textit{is} a `thing' and from whence does it arise? Several approaches to this problem have been proposed.  Of particular interest is the aforementioned work of D\"{o}ring and Isham \cite{Doring:2011fk} and that of Wheeler \cite{Wheeler:1990uq} himself.  Both question what is seemingly one of the unassailably solid pillars upon which modern science is constructed: real numbers and, most notably, their continuity as \emph{physical} fact.  D\"{o}ring and Isham have suggested that physical quantities need not be real-valued \cite{Doring:2011fk}.  Their argument against the supposition of real-valuedness for physically measurable quantities is partly based on a line of reasoning that connects measuring devices with continuous, smooth manifolds and equates such manifolds with `classicality.'  Wheeler is more blunt; he flatly states ``no continuum''~\cite{Wheeler:1990uq}.

The concept of a discrete physical reality is a very old idea. The early atomists of ancient India and Greece theorized that nature consisted of two fundamental concepts: \textit{atom}, which was the presence of something, and \textit{void} which was the presence of nothing. The former represented physical reality in its most basic form. In the context of modern physics, these concepts have a deep relation to the dual notions of space and time: anything that is physically real \textit{occupies} space and time, while the complete absence of any such occupation corresponds to a vacuum\footnote{It is worth noting that in quantum theory the vacuum may be represented by a quantum state $\ket{vac}$. This would seem to blur the distinction between `being' and `nothingness,' but we will leave that discussion for another time.}. In order to `occupy' space and time, a physical `thing' must generally possess measurable properties or characteristics that provide a means by which that `occupation' may be measured. As philosopher Eugene Gendlin has noted, Heidegger's notion of `thing' is really an explanatory approach that ``renders whatever we study as some thing in space, located over there, subsisting separate from \ldots us''~\cite{Gendlin:1967fk}. In other words, a `thing' occupies space and time.

Feynman took the position that anything that possesses energy and momentum\footnote{We will not concern ourselves in this essay with the nature of momentum and energy.} is physically real, i.e. particles and fields\footnote{Quantum field theory has rendered the difference between ÔparticleÕ and ÔfieldÕ virtually meaningless: a particle \textit{is} the quantization of a field.}~\cite{Feynman:1963zr}. More generally, Eddington viewed particles and fields as carriers of sets of ``variates''~\cite{Eddington:1939fk}. Mathematically, such variates are manifest in symmetries that 
represent degrees of freedom of the overall state space of a particle or field with each symmetry making up a sub-space. These symmetries, very generally, provide various means by which particles and fields, along with their configurations and interactions, may be distinguished from one another. For the purposes of this essay, I shall refer generally to anything that `occupies' space and time as \textit{matter-energy}.

Distinguishability, of course, is at the heart of information. As Schumacher and Westmoreland note, ``\textit{Information} is the ability to distinguish reliably between possible alternatives''~\cite{Schumacher:2010uq}. In this sense, information is encoded in the properties of the particles. Or, as Wheeler saw it, the act of distinguishing one alternative from another actually gives rise to the particles themselves, hence his use of the term ``participatory universe.'' His approach began with the working hypothesis that
\begin{quote}
every \textit{it}---every particle, every field of force, even the spacetime continuum itself---derives its function, its meaning, its very existence entirely---even if in some contexts indirectly---from the apparatus-elicited answers to yes-or-no questions, binary choices, \textit{bits}.~\cite{Wheeler:1990uq}
\end{quote}

D\"{o}ring and Isham formalize this in the notion of a \textit{topos} which is a type of mathematical structure known as a \textit{category}. One way to think of a category is as a set of objects that has some connective pattern between the objects~\cite{Spivak:2013fk}. Another way to describe a category is as a mathematical structure consisting of objects and arrows~\cite{Awodey:2010uq}. A topos contains two special objects: a \textit{state object} $s$, and a \textit{quantity-value object} $v$. A given physical quantity $q$ is represented by an arrow $q:s\to v$ in the topos. As D\"{o}ring and Isham note, ``[w]hatever meaning can be ascribed to the concept of the `value' of a physical quantity is encoded in (or derived from) this representation''~\cite{Doring:2011fk}. In this way a `thing' is then somewhat loosely defined as a bundle of properties wherein these properties refer to values of physical quantities. This is more abstract than Eddington's view as it is not at all clear from this whether it necessarily implies an occupation of space and time. From Wheeler's perspective, these quantities represented the answers to questions we put to nature. At the most fundamental level, he believed that all such questions could be reduced to those for which `yes' or `no' were the only possible answers and would thus be represented by a binary digit, i.e. a \textit{bit}\footnote{The first use of the word `bit' in the sense of a binary digit was in Claude Shannon's seminal 1948 paper on information theory in which he ascribed the origin of the term to John Tukey who had written a memo on which the term `binary digit' had been contracted to `bit'~\cite{Shannon:1948kx}.}. Hence, \textit{it from bit}, according to Wheeler, and thus all physical `things' are ultimately information-theoretic in origin.

A more formal, ``rigorously qualitative'' definition of information can be given in terms of the order on a \textit{domain}~\cite{Martin:2011fk}. A domain $(D,\sqsubseteq)$ is a set of objects $D$ together with a partial order $\sqsubseteq$ that includes certain intrinsic notions of completeness and approximation that are defined by this order. For instance, consider two objects $x,y\in D$. The statement $x \sqsubseteq y$ essentially says that $x$ contains (or carries) some (possibly all) information about $y$., i.e. $y$ is ``more informative'' than $x$~\cite{Martin:2011fk}. For example, a Honus Wagner baseball card contains information about Honus Wagner~\footnote{Famously, the T206 Honus Wagner card, distributed between 1909 and 1911, is the most expensive trading card in history, one having sold in 2007 for \$2.8 million.}. Clearly Honus Wagner himself would be far more informative about his life than his baseball card. In the event that $x$ \textit{does} contain the full information about $y$, then $x=y$ and $x$ is said to be a \textit{maximal element} (object) of the domain, in which case it is an example of an \textit{ideal element}. An object that is not ideal is said to be \textit{partial}. So, given a domain that includes both Honus Wagner and his baseball card, Honus Wagner would be a maximal element while his baseball card would be partial.  

A measurement is then understood as a particular type of mapping on a domain that formalizes the notion of \textit{information content}. Specifically, for a map to be considered a measurement and thus a measure of information content, it must, \textit{at a minimum}, be able to distinguish between those elements that it claims are maximally informative (recall the definition of information given by Schumacher and Westmoreland). The details of the formalism are beyond the scope of this essay, but the main point is that the formalism implies the existence of a purely structural relationship between two different classes of informative objects, neither of which need consist of numbers. Formally, this relationship is given by the map $\mu : D\to E$ where it is said that $\mu$ \textit{reflects} properties of simpler objects $E$ onto more complex objects $D$~\cite{Coecke:2011uq}. For instance, the set of characteristics that describe Honus Wagner, say as they appear on his baseball card, would be elements of $E$ whereas the \textit{actual} set of characteristics that `are' Honus Wagner would be elements of $D$. The act of looking at the baseball card is then a measurement $\mu$ and amounts to \textit{inferring} (i.e. `reflecting') something about Honus Wagner by looking at his baseball card. In a sense, then, $\mu$ is just a function that assigns a `value' to each ``informative object'' on a domain that measures its amount of partiality where we understand that `value' does not necessarily have to mean `a number' (e.g. it could be the portrait of Honus Wagner that appears on his baseball card)~\cite{Martin:2011fk,Coecke:2011uq}. Given that a physical quantity $q$ really has no meaning outside the act of measurement $\mu$  we can view $q$ as a specific instantiation or representation of $\mu$ where $s\in D$ ($s$ is an element of $D$) and $v\in E$ ($v$ is an element of $E$). Combining this formalism with Wheeler's assertion that the answers to all \textit{fundamental} measurement `questions' are binary, a `bit' is best understood as a quantity-value object $v$ while the numerical result embodied by the bit is $q\in\{0,1\}$ (i.e. the value of $q$ comes from the set of numbers consisting of 0 and 1). 

This is a \textit{crucial} distinction: \textbf{a bit is \textit{not} the same thing as its value}. To see why, suppose we were presented with the result of a particular measurement and the numerical value of that result was 1. Further suppose that this is all the information we have about the measurement. We cannot, with certainty, say that 1 is the value of a bit since it could equally well be the value of a `trit' (that is anything that may exist in one of only \textit{three} mutually exclusive states) or any other `-it', for that matter. In order for us to know that we have been given the value of a \textit{bit}, we must know that the domain of the quantity-value object is $[0,1]\in\mathbbm{N}$ where $\mathbbm{N}$ is the set of all natural numbers\footnote{Again, this notation is meant to formalize the notion that the only values that $q$ may take are 0 and 1.}. Hence, a bit (and any other `-it') is really a \textit{domain}.
\begin{definition}[Definition 1 (Bit)]
A `bit' is any instantiation of the domain $[0,1]\in\mathbbm{N}$ in which the values of the domain represent mutually exclusive\footnote{The requirement of mutual exclusivity is used to distinguish a `bit' from a `qubit' where the latter allows for superpositions of 0 and 1.} states.
\end{definition}
Information, as defined by Schumacher and Westmoreland, may then be quantified by the probability $P_{S}(q)$ of successfully determining $q$.

\section{Information content}
Eddington referred to knowledge obtained from the act of measurement as \textit{a posteriori} knowledge~\cite{Eddington:1939fk}. Any fore-knowledge of a system prior to such an act is \textit{a priori}. For example, prior to opening a pack of baseball trading cards, we fully expect that the pack will only contain \textit{baseball} trading cards as opposed to trading cards for some other sport. Once the pack has been opened and the cards examined, we now know exactly \textit{which} cards are contained within. The knowledge that the pack contains only \textit{baseball} cards is \textit{a priori} whereas the knowledge that the pack, for example, contains a Honus Wagner card is \textit{a posteriori}.

Technically, Eddington referred to \textit{a priori} knowledge as any knowledge that is derived from a study of the actual \textit{procedure} of measurement~\cite{Eddington:1939fk}. In that sense, baseball cards are less illustrative than quantum states. Consider the quantum state
$$
\ket{\psi_{a^{\prime}}}=\sum_{a^{\prime}}c_{a^{\prime}}\ket{a^{\prime}}
$$
where $c_{a^{\prime}}$ is a complex number and $\ket{a^{\prime}}$ represents a set of basis vectors for some spin axis $a^{\prime}$. The state, $\ket{\psi_{a^{\prime}}}$, represents a state of \emph{a priori} knowledge about an element of the universe since it is not the result of a measurement but rather some observation about the system and what sorts of measurements on that system are possible. This is analogous to knowing that a pack of trading cards contains \textit{baseball} cards, but not knowing which specific cards it contains.

Now suppose that we perform a measurement, $\mathbf{S}_{z}$, on this state such that $
\ket{a^{\prime}}=\mathbf{S}_{z}\ket{\psi_{a^{\prime}}}=+\frac{\hbar}{2}\ket{z+}$. In other words, we are supposing that our measurement of the spin along the $z$-axis yields a value of $+\hbar/2$ with certainty. In this case, the state $\ket{a^{\prime}}$ represents a state of \emph{a posteriori} knowledge about an element of the universe because it provides information about the \textit{actual} state of the system, not just the \textit{possible} state or states of the system.

What is the origin of \textit{a priori} knowledge? How do we know that $\ket{a^{\prime}}$ represents a spin state as opposed to, say, an energy state or momentum state? Analogously, how do we know that a pack of trading cards specifically contains \textit{baseball} cards? In the latter (and purely classical) case, it is clear that some sort of `measurement' (in a loose sense) had to have taken place, i.e. the package presumably has some identifying characteristics on it in order to differentiate it from other types of trading cards. Reading the package essentially constitutes an act of measurement. Thus there is really a \textit{sequence} of processes that leads to maximal knowledge of a system. If we start with a complete lack of knowledge such that our sequence of measurements could lead us to literally \textit{any} final result --- a Honus Wagner baseball card, a fifty-seven-year-old elephant, or a sixteen-inch diameter pizza --- each measurement reduces the range of possibilities from a nearly infinite number down to just one in the end. Thus, every time we make a measurement, we further refine our knowledge of the system, increasing the amount of information we have collected and decreasing the amount of information that we lack. 

We quantify the \textit{lack} of information via statistical entropy since it is zero when the state is exactly known. Specifically, given an object $x\in D$ and a measurement $\mu : D\to E$, the Shannon entropy is given as~\cite{Martin:2011fk,Coecke:2011uq}
$$
\mu x =-\sum_{i=1}^{n}x_{i}\log x_{i} \quad \textrm{with} \quad x \sqsubseteq y \;\Rightarrow\; \mu x \ge \mu y
$$
where for some value $n=N$, $x=y$ and $\mu x = \mu y$. So if $\mu : D \to E$ is a measurement and $x$ is an object that it measures, then the mathematical statement $\mu x\in\textrm{max}(E)\;\Rightarrow\; x\in\textrm{max}(D)$ says that when $\mu x$ reaches its maximum value, then we have obtained as much information as we can about $x$. Thus, if $s\equiv y$ and $v\equiv x$, as our knowledge of the value object $v$ increases, it approaches the maximal element (the state object $s$) which is mathematically written $v\to s$. Simultaneously the entropy decreases such that $\mu v \to \mu s$ and $\mu$ is said to be \textit{monotone}. Note that, whereas entropy is a measure of information \textit{content} and probability, in the manner described above, is a measure of information, subject to a few `moderate' hypotheses, information behaves in the same manner as its content~\cite{Coecke:2011uq}.

As an example, consider a jigsaw puzzle that contains a binary message that is only decipherable when the puzzle has been completed where on each puzzle piece is printed the value of exactly one bit (i.e. a 0 or a 1). We may represent the state of the puzzle's message as $s$. Each time we place a puzzle piece represented as $v_{i}$, the partially completed puzzle represents a different value object, $v$ and we gain one new bit, $q_{n}$, of information. When we have placed the final piece, $n=N$ in which case we have obtained maximal knowledge of the message and $\mu v$ will have reached a minimum (in fact it should be zero unless we are missing a piece of the puzzle).

In classical physics, we typically assume that if we are armed with \textit{a priori} knowledge about a system, all \textit{a posteriori} knowledge about that system may be inferred. In other words, in such a system, while $N$ may (or may not) be infinite, there may a threshold $v_{\textrm{min}}$ such that if $v_{\textrm{min}}\le v$, $s$ may be predicted with \textit{near} certainty. For example, it may be that we can accurately predict the puzzle's message with only some fraction of the pieces having been assembled. As another example, suppose that the information about the state of a system is fully encoded in the fraction $\frac{5}{6}=0.8\bar{3}$. Clearly knowledge of just one significant digit is not enough information to predict the state with anywhere near certainty since, for example, $0.8=\frac{4}{5}$. While perfect certainty in this example is impossible since $\frac{5}{6}$ is a non-terminating decimal fraction, we can at least establish a limit such that, at some point, we may say with confidence that $v\approx s$ (i.e. at some point we can be fairly certain that the state is $\frac{5}{6}$.

A universe whose future states may be predicted with certainty based on a complete knowledge of its prior states may be said to be \textit{physically deterministic}. We may phrase the condition corresponding to physical determinism in a more rigorous and mathematical manner in terms of state objects and quantity-value objects as follows.
\begin{definition}[Condition 1 (Physical determinism)]
Let $u\equiv s$ be the ideal element on the domain of physical measurements that provide information about the universe. Then, if $u$ is static and either $N$ is finite or $u$ is predictable then, $v_{\textrm{min}}\le v \;\Rightarrow\; v\to u.$
\end{definition}
A hypothetically omniscient being who happens to be in possession of $q_{\textrm{min}}\le q_{n}$ bits, where $q_{\textrm{min}}:u\to v_{\textrm{min}}$ (i.e. possesses enough bits of information about the universe to fully predict its future states), is known as Laplace's demon. 

One of the assumptions of physical determinism is that some properties are considered to be immutable --- once a Honus Wagner baseball card, always a Honus Wagner baseball card. This is not without its problems as it implies that $\mu v\to \mu u$, i.e. as we obtain more and more information about the universe, its entropy should \textit{decrease}. The second law of thermodynamics tells us, of course, that the exact opposite is actually happening and as Eddington famously said, ``if your theory is found to be against the second law of thermodynamics I can give you no hope; there is nothing for it but to collapse in deepest humiliation''~\cite{Eddington:1928fk}. Something is clearly amiss.

The problem lies in the seemingly innocuous assumption that by increasing our knowledge of a physical system we will necessarily arrive at a full description of the system, i.e. that, as $n\to N$, it must be that $v\to s$. For this to be the case, $s$ would have to be determinable. Consider once again the jigsaw puzzle containing a binary message. For $s$ to be determinable, it must be \textit{static} (e.g. the message cannot change as we assemble the puzzle since, if it did, the pieces would need to be reorganized) \textit{and} either $N$ would have to be finite (e.g. the puzzle would have to contain a finite number of pieces) or $s$ would have to be predictable in some manner (e.g. the message would have to have a predictable pattern). While in some cases $s$ may not be determinable with perfect certainty, as I noted previously, in some cases we can establish a limit whereby we may say with some degree of confidence that $v\approx s$. This is frequently the case in classical systems. To use an old adage, ``close only counts in horseshoes and hand grenades'' --- and classical physics. Of course things become a bit more difficult at the quantum level.

\section{Contextuality}
Given two objects $x,y \in D$, the statement $x\preceq y$ is read ``$x$ \textit{approximates} $y$''~\footnote{The notation $\ll$ is standard but, given the more general audience of this essay, I have adopted $\preceq$ so as to clearly distinguish it from the usual meaning of $\ll$ in inequalities.}. It means that $x$ carries some \textit{essential} information about $y$ where we can think of `essential' as being synonymous with `indispensable.' In other words, while $x$ may not carry \textit{all} the information about $y$, it carries information that is \textit{necessary}. For instance, in the jigsaw puzzle example, there may be certain bits of the message that are indispensable in order for it to be read or comprehended. Any piece that has the value of an indispensable bit printed on it would then be essential.

There is an inherent context embedded in the statement $x\preceq y$. Consider three objects, $x,y,z \in D$ and suppose that $x\preceq y$. Also suppose that $y \sqsubseteq z$. This means that $x$ carries \textit{essential} information about $y$ and $y$ carries some (not necessarily essential) information about $z$. In order for us to conclude from this that $x\preceq z$, we would need to know that the statement $y \sqsubseteq z$ is being made in the same \textit{context} as $x \preceq y$. We understand `context' to mean a `setting' within which we make a statement. The results of classical measurements are elements of \textit{continuous domains}. This simply means that classical variables that correspond to measurements may take a continuous (as opposed to discrete) range of values. Approximation on continuous domains is \textit{context independent}~\cite{Coecke:2011uq}. This means that for classical measurements, it is automatically true that if $x\preceq y$ and $y \sqsubseteq z$, then $x\preceq z$. So, for instance, if a certain piece of our aforementioned jigsaw puzzle is essential, it is essential regardless of how or where (or even when) we assemble the puzzle. Given the limitations on space, I will refer those interested in a more basic explanation of contextuality to Ref.~\cite{Bacon:2008uq}.

\begin{figure}
\begin{center}
\begin{tikzpicture}
\node at (-1.5,-0.1)[left] {$\ket{\psi}$};
\draw[ultra thick,gray!30!yellow!60!orange,-latex'] (-1.5,-0.1) -- (-0.9,-0.1);
\draw[ultra thick,gray!30!yellow!60!orange] (-1,-0.1) -- (-0.6,-0.1);
\node at (0,0) {
  \begin{tikzpicture}
  \shade[left color=gray!50!white,right color=blue!25!white!25!gray] (0,1) rectangle (1,0);
  \shade[top color=gray!50!white,bottom color=blue!25!white!25!gray](0,1) -- (0.25,1.25) -- (1.25,1.25) -- (1,1) -- cycle;
  \shade[top color=gray!50!white,bottom color=blue!25!white!25!gray] (1.25,1.25) -- (1.25,0.25) -- (1,0) -- (1,1) -- cycle;
  \fill[black!50!gray] (1.125,0.85) ellipse (0.0525cm and 0.15cm);
  \fill[black!50!gray] (1.125,0.4) ellipse (0.0525cm and 0.15cm);
  \fill[red] (0.6,1.125) arc (0:180:0.15cm and 0.15cm);
  \fill[blue] (1,1.125) arc (0:180:0.15cm and 0.15cm);
  \end{tikzpicture}};
\node at (0,-1) {$\mathbf{S}_{a}$};
\draw[ultra thick,gray!30!yellow!60!orange,-latex'] (0.475,0.2) -- (1.1,0.2);
\draw[ultra thick,gray!30!yellow!60!orange] (1,0.2) -- (1.5,0.2);
\node at (2,0.25) {
  \begin{tikzpicture}
  \shade[left color=gray!50!white,right color=blue!25!white!25!gray] (0,1) rectangle (1,0);
  \shade[top color=gray!50!white,bottom color=blue!25!white!25!gray](0,1) -- (0.25,1.25) -- (1.25,1.25) -- (1,1) -- cycle;
  \shade[top color=gray!50!white,bottom color=blue!25!white!25!gray] (1.25,1.25) -- (1.25,0.25) -- (1,0) -- (1,1) -- cycle;
  \fill[black!50!gray] (1.125,0.85) ellipse (0.0525cm and 0.15cm);
  \fill[black!50!gray] (1.125,0.4) ellipse (0.0525cm and 0.15cm);
  \fill[red] (0.6,1.125) arc (0:180:0.15cm and 0.15cm);
  \fill[blue] (1,1.125) arc (0:180:0.15cm and 0.15cm);
  \end{tikzpicture}};
\node at (2,-1) {$\mathbf{S}_{b}$};
\draw[ultra thick,gray!30!yellow!60!orange,-latex'] (2.475,0) -- (3.1,0);
\draw[ultra thick,gray!30!yellow!60!orange] (3,0) -- (3.5,0);
\node at (4,0.1) {
  \begin{tikzpicture}
  \shade[left color=gray!50!white,right color=blue!25!white!25!gray] (0,1) rectangle (1,0);
  \shade[top color=gray!50!white,bottom color=blue!25!white!25!gray](0,1) -- (0.25,1.25) -- (1.25,1.25) -- (1,1) -- cycle;
  \shade[top color=gray!50!white,bottom color=blue!25!white!25!gray] (1.25,1.25) -- (1.25,0.25) -- (1,0) -- (1,1) -- cycle;
  \fill[black!50!gray] (1.125,0.85) ellipse (0.0525cm and 0.15cm);
  \fill[black!50!gray] (1.125,0.4) ellipse (0.0525cm and 0.15cm);
  \fill[red] (0.6,1.125) arc (0:180:0.15cm and 0.15cm);
  \fill[blue] (1,1.125) arc (0:180:0.15cm and 0.15cm);
  \end{tikzpicture}};
\node at (4,-1) {$\mathbf{S}_{c}$};
\draw[ultra thick,gray!30!yellow!60!orange,-latex'] (4.475,0.31) -- (5.25,0.31);
\draw[ultra thick,gray!30!yellow!60!orange,-latex'] (4.475,-0.14) -- (5.25,-0.14);
\node at (5.25,0.31)[right] {$+$};
\node at (5.25,-0.14)[right] {$-$};
\node at (5.75,0.085)[right] {\Large{\}} \normalsize{?}};
\end{tikzpicture}
\end{center}
\caption{\label{fig1} Each box represents a measurement of the spin for a spin-$\frac{1}{2}$ particle along some axis with the top output indicating that the state is \textit{aligned} ($+$) with the measurement axis, and the bottom output indicating that the state is \textit{anti-aligned} ($-$) with the measurement axis. Red and blue lights on the top simply indicate to the experimenter which of the two results is obtained (e.g. red might indicate aligned and blue might indicate anti-aligned).}
\end{figure}
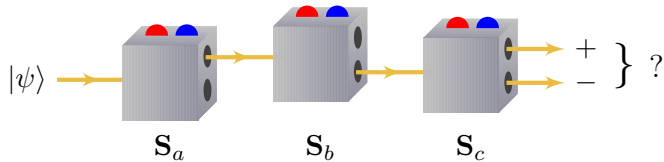

In quantum systems a `context' is related to a measurement basis\footnote{We point those readers interested in a refresher on measurements and bases in quantum mechanics to Ref.~\cite{Schumacher:2010uq}.}. So mathematically one way to describe a context is as a domain $(\Omega[\mathsf{m}],\sqsubseteq )$ where for $v_{i},s \in \Omega[\mathsf{m}]$ and $v_{i}\sqsubseteq s$, a specific measurement yields $q_{i}[\mathsf{m}]: s[\mathsf{m}]\to v_{i}[\mathsf{m}]$ and $\mathsf{m}$ identifies the measurement basis. For an orthonormal $n$-dimensional basis, we may write $v_{1}[\mathsf{m}]\perp v_{2}[\mathsf{m}] \perp \cdots \perp v_{n}[\mathsf{m}]$ where the symbol $\perp$ is used to indicate the fact that any two $v_{i}$ in such a basis represent mutually exclusive results. For example, in a two dimensional basis that yields measurement values $q_{1}$ and $q_{2}$, we have $q_{1}: s \to v_{1}$ and $q_{2}: s \to v_{2}$ where $v_{1} \perp v_{2}$. The values $q_{1}$ and $q_{2}$ might be $0$ and $1$ or they might be $+1$ and $-1$ or even some other set of values entirely ---  they quite literally correspond to something we `read' off a device which means it doesn't even have to be a number at all (see Fig.~\ref{fig1} for example). The key is that the quantity-value objects represent the more abstract elements of the basis and thus are where the orthonormality manifests itself. Thus orthogonality provides an order-theoretic definition of what it means for two `objects' (elements, states, measurements, etc.) to truly be \textit{distinct}. Any two objects that are not orthogonal must share something in common. A more detailed mathematical treatment may be found in Appendix A: Technical End Notes.

Now consider a sequence of three spin-measurement devices that measure along axes $a$, $b$, and $c$. Note that these axes themselves do not necessarily need to be in any way perpendicular to one another. Each represents a \textit{separate} orthonormal basis. Suppose, then, that the state of the particle exiting the second device is $\ket{b-}$, as shown in Fig.~\ref{fig1}.
Standard quantum mechanics tells us that the associated probabilities for the results of the third measurement are $P_{c}(+) = \sin^{2}\frac{1}{2}\theta_{bc}$ and $P_{c}(-) = \cos^{2}\frac{1}{2}\theta_{bc}$ where $\theta_{bc}$ is the angle between the $b$ and $c$ axes. So, for example, if $\theta_{bc}=90^{\circ}$ then $P_{c}(+)=P_{c}(-)=0.5$ which is the same result we would find if the state of the particle exiting the second device was completely randomly oriented. In other words, this predicts a perfectly random measurement result and means we can't obtain any useful information about axis $b$ via this measurement. Mathematically we can quantify this by writing $s[b]\cap s[c]=0$ which means that the state objects for the particle in each instance share nothing in common. Conversely, if $\theta_{bc}=0^{\circ}$ which means that $b=c$, then it is as if we are merely confirming the second measurement: we are guaranteed to find that the particle is in the state $\ket{c-}$. This, of course, is the exact opposite of perfect randomness and so we can write $s[b]\cap s[c]=1$ corresponding to full knowledge of axis $b$.

In a sense, then, the statement $s[\mathsf{m}]\cap s[\mathsf{n}]$ \textit{quantifies} contextuality and is a bit like a dot product between two vectors with one crucial difference. Suppose that $\theta_{ab}=\theta_{bc}=90^{\circ}$ and that $a=c$, i.e. they represent the same axis. Let us suppose that the state subsequent to the first measurement is, as in Fig.~\ref{fig1}, $\ket{a+}$ and that the state subsequent to the second measurement is, also as in Fig.~\ref{fig1}, $\ket{b-}$ (in fact it doesn't really matter if it is $\ket{b-}$ or $\ket{b+}$ as long as the angles are as described). We already know that the probabilities for the results of the third measurement are $P_{c}(+)=P_{c}(-)=0.5$. This means that it is entirely possible for the state subsequent to the third measurement to be $\ket{c-}=\ket{a-}$. Clearly $s[a]\cap s[c]=0$  \textit{even though a and c are the same axis}! Thus contextuality provides a means by which a quantum state can essentially be `reset.' Alternatively one could say that the particle has no `memory' of having been in the $\ket{a+}$ state.

Note that though the actual result of a measurement itself could be entirely random, the basis is chosen in a purely deterministic manner i.e. we never find a measurement result in a basis \textit{other} than the one in which we choose to make the measurement. For example, if we open a pack of baseball cards we know we won't find that it contains playing cards (or, in even simpler terms, you never find an orange growing on an apple tree\footnote{If you ever do, run like hell. The zombies are coming.}).

Quantum states are, of course, more complicated than classical states: they may be represented by a density matrix as opposed to a simple number and the entries of the density matrix may be complex. Crucially, however, the result of a quantum measurement will always produce a result that  lies in the domain of \textit{classical} quantity-value objects. For example, though a qubit may exist in some mixture of $\ket{0}$ and $\ket{1}$, when measured it is always found to be in \textit{either} $\ket{0}$ \textit{or} $\ket{1}$.  The key here is that quantum states may be \textit{informationally isolated} in which case they may exist in a superposition or a mixed state. Once a measurement is made, however, they are no longer informationally isolated and any record of the superposition or mixture is lost~\cite{Schumacher:2010uq}. To put it another way, the domain of classical states $D$ is \textit{strictly} smaller than the domain of quantum states $\Omega$. I will write this $D \sqsubset \Omega$ where I take $D \sqsubset \Omega$ to mean ``D carries some, but strictly \textit{not} all, information about $\Omega$''\footnote{This is non-standard notation that I introduce here for the sake of simplifying the presentation.}. Thus a quantum measurement is a map $q: \Omega \to D$ and the von Neumann entropy
$$
\sigma\rho=-\mathrm{tr}(\rho\log\rho)
$$
on quantum states is thus also a measurement $\sigma : \Omega \to E$ where $\sigma$ factors as $\sigma = \mu \circ q$ and $\mu : D \to E$ is a classical measurement~\cite{Coecke:2011uq}. As the example given in Fig.~\ref{fig1} demonstrates, the loss of informational isolation is intimately related to contextuality. Thus contextuality provides for two important features in quantum states: they may be `reset' and it is possible for them to store more information than can be effectively obtained via measurement.

These features help to solve the mystery we noted before: if we are constantly learning more and more about the universe (via measurements), why is the universe's entropy increasing? The first feature explains why the universe's entropy is not \textit{decreasing} by telling us that, rather than gaining knowledge about the universe with each new measurement we make, we might only hope to `break even' since we essentially have to start over (in a way) with each new measurement. The second feature, however, when combined with the first guarantees that this is a false hope because every time the reset occurs, more information (in the form of a superposition or mixture) about the universe is created that is then lost in a measurement. The only way to prevent this is for the state to remain unchanged as I noted earlier which is only possible if we \textit{never change our measurement basis}. But in the wonderful world of quantum mechanics, any transformation or interaction is essentially the same as a measurement. Hence, contextuality directly leads to an ever-increasing entropy for the universe, i.e. the second law of thermodynamics is a direct consequence of quantum contextuality\footnote{I suppose it is possible to take the opposite view that the second law leads to the requirement of contextuality, but that is really a `chicken and egg' type of argument.} \textit{and}, because $u$ is not static, the universe as a whole must \textit{not} be physically deterministic!

\section{It or bit?}
These considerations then bring us back to Wheeler's original declaration that every `it' derives its very existence from `bits' of information. But while the information content of the universe is constantly increasing with each new `measurement' (interaction), the matter-energy content of the universe is known to be constant. It would seem that if Wheeler were literally correct, this latter point should not be true. In other words, if `it' truly --- literally --- comes from `bit' and the number of bits of information in the universe is always increasing, why doesn't this result in the creation of at least \textit{some} new matter-energy? In other words, even if not \textit{every} new bit of information necessarily led to some new `it,' it seems reasonable to assume that at least \textit{some} would.

In fact there is really nothing mysterious about the ever-increasing entropy if one takes the Schumacher and Westmoreland definition of information literally as ``the ability to distinguish reliably between possible alternatives''~\cite{Schumacher:2010uq}. In this sense, entropy is merely a measure of the number of possible configurations of the system and contextuality guarantees that we are provided with an ever-increasing number of them. Thus it seems quite logical to conclude the exact opposite of Wheeler --- `bit' actually comes from `it' and is really a direct consequence of quantum contextuality since it is contextuality that creates the alternatives (and hence the need to distinguish between them) in the first place. Ironically, because contextuality also implies that the universe is not physically deterministic, perhaps it satisfies Wheeler's notion of a universal regulating principle arising from a ``regularity based on chaos, of `law without law' '' \cite{Wheeler:1983fk}. Indeed, perhaps the key to understanding the universe is lying right under our very noses.

\newpage

\bibliographystyle{plain}
\bibliography{FQXi4.bib}

\newpage

\appendix

\section{Technical end notes}
Orthogonality plays such an important role here, as evidenced by the example of Fig.~\ref{fig1}, that it warrants a more detailed treatment. 

Martin (~\cite{Martin:2011fk}) gives the following definition of a \textit{dcpo} which is a particular type of domain that is integral to the work presented here.
\begin{definition}[Definition 2 (dcpo)] Let $(P,\sqsubseteq)$ be a partially ordered set or \textit{poset}. A nonempty subset $S\subseteq P$ is \textit{directed} if $(\forall x,y\in S)(\exists z \in S)x,y\sqsubseteq z$. The \textit{supremum} $\bigsqcup S$ of $S\subseteq P$ is the least of its upper bounds when it exists. A \textit{dcpo} is a poset in which every directed set has a supremum.
\end{definition}
Any continuous dcpo is an example of a domain. Now let $(D,\sqsubseteq)$ be a dcpo. For elements $x,y \in D$ we set
$$
\uparrow x := \{y\in D: x\sqsubseteq y\} \quad \textrm{and} \quad \downarrow x:= \{y\in D: y\sqsubseteq x\}.
$$
Then for some dcpo $D$, a pair of elements $x,y \in D$ are said to be \textit{orthogonal} if $$\mu(\uparrow x\;\cap\uparrow y)\subseteq \{0\}$$ which may be written $x\perp y$.

\end{document}